\documentclass[
]{ceurart}

\usepackage{listings}
\usepackage[ruled,linesnumbered]{algorithm2e}

\usepackage{tikz}
\usetikzlibrary{arrows.meta, positioning, shapes.geometric}
\usepackage{graphicx}
\usepackage{subcaption}
\usepackage[ruled,linesnumbered]{algorithm2e}
\usepackage[most]{tcolorbox}
\tcbset{
  colback=gray!5!white, 
  colframe=gray!50!black, 
  colbacktitle=black,
  coltitle=white,
  fonttitle=\bfseries\footnotesize,
  boxrule=0.5pt, 
  arc=3pt, 
  left=6pt, 
  right=6pt, 
  top=2pt, 
  bottom=2pt, 
  boxsep=4pt,
  enhanced,
  breakable,
}
\usepackage{listings}
\lstdefinestyle{codeblock}{
    basicstyle=\ttfamily\small,
    keywordstyle=\color{blue!70!black},
    commentstyle=\color{gray!80},
    stringstyle=\color{green!50!black},
    showstringspaces=false,
    breaklines=true,
    frame=single,
    framerule=0.2pt,
    backgroundcolor=\color{gray!5}
}
\usepackage{subcaption}
\usepackage{cleveref}

\begin{document}
\copyrightyear{2026}
\copyrightclause{Copyright for this paper by its authors.
  Use permitted under Creative Commons License Attribution 4.0
  International (CC BY 4.0).}

\conference{LLM-Text2KG'26: 5th International Workshop on LLM-Integrated Knowledge Graph Generation from Text (Co-located with ESWC 2026),
  May 10--14, 2026, Dubrovnik, Croatia}

\title{From ``Strings'' to ``Things'' for Personal
Knowledge Graphs: Evaluating LLM Triple
Extraction for Recommendation Systems}

\author{Abhirup Dasgupta}[%
orcid=0009-0002-8434-9754
]
\cormark[1]
\fnmark[1]
\address{Rensselaer Polytechnic Institute, Troy NY 12180, USA}

\author{Fernando Spadea}[%
orcid=0009-0006-4278-3666
]
\cormark[1]
\fnmark[1]

\author{Oshani Seneviratne}[%
orcid=0000-0001-8518-917X
]
\cormark[1]

\cortext[1]{Corresponding author.}
\fntext[1]{These authors contributed equally.}

\begin{abstract}
Personal Knowledge Graphs (PKGs) offer a privacy-preserving framework for modeling user preferences, yet constructing them from unstructured, decentralized conversational data remains a challenge. This paper bridges the gap between conversational ``strings'' and semantic ``things'' by presenting a reproducible pipeline for extracting structured user-preference triples using lightweight Large Language Models (LLMs). We evaluate Qwen- and Gemma-based models on their ability to extract RDF-compliant triples linked to Wikidata identifiers from conversational data for PKG construction.
Our evaluation assesses both the semantic extraction fidelity and the utility of the resulting graphs in a downstream recommendation task.
We found that certain models performed well and had proportionally high downstream performance relative to their triple extraction performance.
\end{abstract}


\maketitle

\section{Introduction}
Personalized Knowledge Graphs (PKGs) have emerged as a powerful paradigm for representing user-specific preferences, contextual signals, and behavioral patterns in a structured and machine-interpretable form. Unlike global knowledge graphs that encode only widely ``notable'' facts, PKGs capture fine-grained, transient, and often non-canonical personal information essential for personalization, user modeling, and adaptive AI systems. Foundational work by Balog and Kenter~\cite{balog2019personal} formally articulated the promise and challenges of PKGs, while subsequent surveys~\cite{skjaeveland2024ecosystem,shirai2020survey} have underscored the need for scalable construction pipelines and interoperable representations. Complementary advances in KG-enhanced recommendation~\cite{wang2019kgat} and conversational preference elicitation~\cite{christakopoulou2016towards} further demonstrate that explicitly structured relational signals consistently improve personalization quality.

Building on this trajectory, we investigate a dialogue-driven PKG construction approach that leverages open-weight large language models (LLMs) to extract user–item relational triples, validate their semantic fidelity, and assess their downstream utility in recommendation settings. In doing so, this work advances creating, managing, and exploiting personalized, dynamically generated KGs that remain interpretable, interoperable, and semantically grounded.

Although pre-trained LLMs excel at open-domain knowledge extraction, their capacity to construct PKGs, particularly from conversational interactions, remains underexplored. Traditional recommender systems that incorporate structured user–item relationships via graph-based neural models have demonstrated clear gains in accuracy and interpretability~\cite{10.1145/3292500.3330989,9270221}. However, such methods typically depend on centrally logged, explicit interaction traces, which are often unavailable due to privacy concerns or infrastructural constraints. Privacy considerations may restrict the collection or sharing of detailed behavioral data, while infrastructural limitations may prevent platforms from maintaining persistent user identifiers, centralized logging pipelines, or cross-service data aggregation. This motivates an alternative paradigm: extracting user preferences directly from natural language dialogue, enabling decentralized PKG construction without persistent tracking or centralized logging.

Entities (users, movies) are assigned Internationalized Resource Identifiers (IRIs), and user–item relations such as \texttt{likes}, \texttt{seen}, and \texttt{suggested} are captured through a small, domain-specific ontology. This yields RDF-compliant graphs that inherit the benefits of formal semantics, global interoperability, and compatibility with downstream reasoning frameworks. We therefore define our workflow as a \emph{Personal Knowledge Graph Construction (PKGC) pipeline} that transforms conversational data into consistently structured, semantically annotated knowledge graphs.
To evaluate the feasibility of this approach, we study the ability of modern LLMs to extract accurate and semantically grounded preference triples from natural multi-turn dialogue in a movie recommendation setting. 
We use the \textit{Recommendation Dialogues} (ReDial) dataset~\cite{li2018towards,huggingface2024redial}, which contains human-to-human conversations paired with structured annotations describing users’ stated or implied preferences. This dataset enables a principled comparison between human-annotated ground truth and LLM-extracted PKG structures.
Our contributions are as follows:
\begin{enumerate}
\item \textbf{A prompt-based PKG construction pipeline} that converts conversational recommendation dialogues into structured PKGs of (User, Relation, Item) triples, using an explicit semantic schema with IRIs and an RDF-compliant relation vocabulary.
\item \textbf{A systematic evaluation of open-weight LLMs} (Qwen3 and Qwen2.5 variants, in multiple sizes and prompting regimes) against the ReDial dataset's structured annotations, providing the first large-scale assessment of conversational PKG extractability.
\item \textbf{An analysis of the downstream utility of extracted PKGs}, demonstrating how extraction fidelity influences recommendation performance and showing that LLM-derived PKGs can serve as a privacy-preserving and interpretable foundation for decentralized personalized services.
\end{enumerate}

\subsection{Motivation}
\label{sec:motivation}
A central motivation for PKGs is their role in enabling personalized recommendations while supporting mechanisms that promote viewpoint diversity, as demonstrated in frameworks such as KG Inversion~\cite{spadea2025bursting}. These approaches assume the availability of a reliable method for constructing PKGs directly from users’ organic interactions, including natural language dialogues. This paper provides that missing foundation by evaluating how effectively LLMs can extract structured preference triples needed for PKG construction.

The core of our approach is to use PKGs as a structured, intermediate representation of user preferences, created specifically to be consumed by an LLM. While an LLM could, in theory, process raw conversational history, such an approach is inefficient for on-device applications. Conversations are often meandering and contain irrelevant information that would needlessly consume the LLM's limited context window. A PKG, by contrast, distills these dialogues into a concise set of salient facts.

One might consider using an LLM to generate a textual summary of the conversation instead. However, this approach presents significant challenges as user history grows. Combining summaries from multiple conversations to create a unified user profile without introducing redundant information would require a sophisticated and computationally expensive LLM-based merging process. In contrast, avoiding redundant information in a PKG is an efficient, algorithmic task; redundant facts (e.g., the same preference expressed in two different conversations) can be trivially ignored or deduplicated. This efficiency is paramount for systems designed to run on user devices.

Furthermore, the structured nature of PKGs makes them highly adaptable and queryable. Retrieving a specific subset of preferences (i.e., all movies a user has seen but dislikes) is a simple graph traversal operation. Achieving the same result from a conversation summary would again require another LLM call to parse the text and extract the relevant information. This adaptability is precisely what enables complex, targeted interventions for promoting recommendation diversity, a task that would be far more complicated and less reliable if performed on unstructured text.

As personalization needs evolve, dynamic PKG adaptation becomes increasingly important. Advances in knowledge editing and continual learning~\cite{meng2022locating,mengmass,de2021editing} offer mechanisms to insert or update specific facts in PKGs before they are presented to an LLM, altering functionality without retraining. These tools could enable PKGs to not only be constructed from conversations but also updated to reflect changing user preferences, a capability missing from static graph models and current dialogue-based extraction pipelines.

Therefore, the central motivation for this research is to evaluate the feasibility of using modern open-weight LLMs for the foundational task of triple extraction. By benchmarking the ability of LLMs to bridge the gap between unstructured dialogue and structured PKGs, we are taking the essential first step toward unlocking a more efficient, adaptable, and privacy-preserving paradigm for personalized recommendations. Additionally, having a downstream recommendation task for our extracted PKGs gives us a domain in which to evaluate the efficacy of the extracted PKGs.

\subsection{Dataset Considerations}
\label{sec:dataset}
To evaluate our approach, we use the ReDial dataset~\cite{li2018towards,huggingface2024redial}, a widely adopted benchmark for conversational recommendation research.
The ReDial dataset consists of 11,348 dialogues between over a thousand users centered on movie recommendations, mirroring the conversational interactions our proposed system is designed to process. A key advantage of ReDial is its inclusion of structured metadata that details user preferences. 

These annotations provide a structured reference that allows us to systematically evaluate the correctness of triples extracted by the LLM-based pipeline.

Another important attribute of ReDial is the conversational diversity it captures. The dataset encompasses a diverse range of conversational styles from over a thousand unique users, featuring varying levels of formality. This diversity ensures the LLMs are benchmarked against realistic, informal language rather than perfectly structured text, a scenario unlikely in real-world recommendation-seeking conversations. 
In this context, ReDial serves as a high-quality testbed, offering both scale and annotation fidelity to rigorously evaluate the performance of LLM-driven PKG extraction pipelines. 

The extraction pipeline itself is not specific to ReDial and can be applied to other conversational datasets or user–assistant interaction logs. In practice, the method only requires dialogues containing identifiable user–item mentions and a schema to map extracted relations into structured triples.

\section{Related Work}

\paragraph{Open-Domain LLM-Based Knowledge Base Construction (KBC).}
LLMs have recently been explored as general-purpose engines for KBC, where early work shows that models such as GPT-3 and T5 can be prompted to produce entity–relation–entity triples from raw text without supervised training~\cite{petroni-etal-2019-language,roberts-etal-2020-much}. More recent prompt- and retrieval-augmented methods~\cite{frey2023benchmarking} refine this capability, improving extraction fidelity while also revealing persistent issues around consistency, hallucination, and coverage. EDC~\cite{Wang2024EDC} proposes a schema-flexible pipeline: Extract, Define, Canonicalize, demonstrating strong performance on open-domain datasets. Multi-agent approaches such as CooperKGC~\cite{Zhang2023CooperKGC} reduce hallucination by cross-validating extractions among LLM agents. Our work is aligned with this literature but differs in scope: rather than constructing general-purpose factual KGs, we specialize the extraction process to detect user preference relations in conversational settings, incorporate domain-specific normalization for movie entities, and ground all extracted triples in a PKG structure.

\paragraph{Dialogue Relation Extraction (DRE).}
This research direction aims to recover relational knowledge from multi-turn conversations. Empirical studies~\cite{li2024empirical} show that frontier LLMs such as ChatGPT and LLaMA-70B perform strongly on DRE benchmarks, benefiting from their ability to reason over long-range dependencies and incomplete context. This suggests LLMs may be well-suited for personalized knowledge mining from conversational data. However, these prior works evaluate extraction against synthetic or task-specific benchmarks, whereas our study conducts a systematic evaluation using human-annotated conversational recommendation data, providing a more realistic and grounded assessment.

\paragraph{Preference Extraction in Recommender Dialogues.}
LLMs have been used both to simulate recommendation dialogues for data augmentation~\cite{liang2024llm} and to extract user preferences from conversation~\cite{kim2025extracting,kook2025empowering}. Benchmarks such as LLM-REDIAL~\cite{liang2024llm} show that off-the-shelf LLMs struggle with next-item prediction in zero- and few-shot settings, with performance improving only when fine-tuned or supplemented with user histories, highlighting the importance of persistent user modeling. Prior work has explored implicit preference detection~\cite{kim2025extracting} and hybrid LLM–retrieval pipelines~\cite{kook2025empowering}, but these systems focus primarily on optimizing downstream recommendation. In contrast, our work centers explicitly on the extraction task itself, quantifying how accurately LLMs can recover user-specific relational triples that align with human annotations.

\paragraph{KG-Enhanced Recommendation Systems.}
KGs are widely used to enrich recommender systems with structured relational context. Prior studies~\cite{Chen2023LKPNR} show that integrating LLM-derived signals into KG-based architectures improves personalization, while industrial systems~\cite{Zhao2024LLMKERec} demonstrate that validating inferred graph edges enhances robustness under sparse data conditions. These works leverage KGs for downstream reasoning and recommendation, whereas our primary contribution lies upstream: we generate the KG itself from conversational interaction and quantify the fidelity of the extracted relations before they are used in recommendation tasks. This distinction positions our method as complementary to KG-grounded recommenders and LLM-based scoring models.

\paragraph{PKGs and User-Centric Semantic Modeling.}
PKGs have been explored as a means of modeling personal, context-rich information. Work in personal health informatics has shown how they support individualized reasoning~\cite{chen2022semantic}, PKG construction from IoT and behavioral data~\cite{seneviratne2023personal}, and personalized diet recommendations grounded in clinical semantics~\cite{seneviratne2021akbcpersonal}. Recent efforts integrate PKGs with causal reasoning and LLMs to generate adaptive, interpretable health insights~\cite{yang2024transforming}, and apply KG-based personalization techniques in nutrition recommendation~\cite{zhang2025adaptive}, and patient modeling~\cite{shirai2021applying}. Our work extends this PKG-centered perspective to the domain of conversational recommendation: instead of deriving PKGs from sensors or structured logs, we study how reliably LLMs can extract personal relational knowledge directly from natural multi-turn dialogue, providing a semantic foundation for decentralized, privacy-preserving personalization.

\section{Conversational Triple Extraction Pipeline}

\begin{figure}
    \centering
    \includegraphics[width=\linewidth]{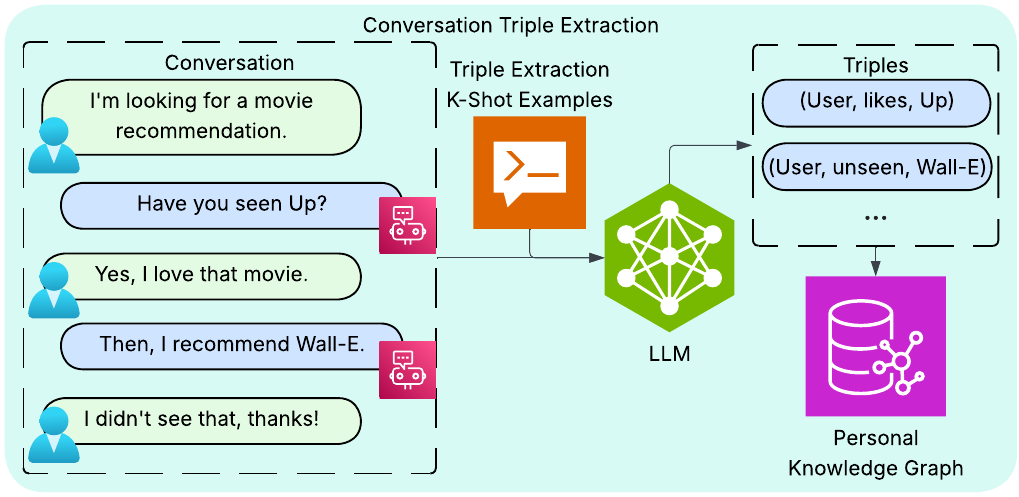}
    \caption{Triple Extraction Pipeline: A natural language conversation between a user and a recommender agent is processed by an LLM to extract structured preference triples (e.g., \texttt{(User, likes, Up)}, \texttt{(User, unseen, Wall-E)}). The model is guided by k-shot examples and predefined relation definitions to ensure consistent output. Extracted triples are stored in a PKG, enabling downstream recommendation and reasoning tasks.}
    \label{fig:pipe}
\end{figure}

Our methodology for the triple extraction pipeline is illustrated in \Cref{fig:pipe}.
The pipeline takes a multi-turn dialogue between a user and a conversational agent as input and produces structured preference triples of the form $(\texttt{User}, \textit{relation}, \texttt{Item})$. 
These triples are subsequently normalized, aligned to external knowledge graph identifiers, and incorporated into a PKG.

For experimental evaluation, we instantiate this pipeline using conversations from the ReDial dataset (\Cref{sec:dataset}). 
ReDial contains dialogues between an initiator and a respondent discussing movie recommendations. 
In our setup, the initiator is treated as the user whose preferences are modeled, while the respondent plays the role of the conversational assistant.

By default, the movies are represented by IDs in the conversations, but we replaced them with their titles instead. 
Each conversation is annotated with metadata features. These features describe the initiator's relationship with each movie mentioned, and we use them to derive relational triples of the form $(\texttt{User}, \textit{relation}, \texttt{Movie})$. The features and the relations they map to are defined in \Cref{tab:feature-mapping}.

\begin{table}
    \centering
    \caption{
    Example mapping between conversational preference signals and PKG relation types (instantiated using ReDial annotations).}
    \label{tab:feature-mapping}
    \begin{tabular}{lll}
        \toprule
        \textbf{Feature} & \textbf{Value} & \textbf{Mapped Relation} \\
        \midrule
        \multirow{3}{*}{\begin{tabular}[c]{@{}c@{}}\texttt{liked} \\ \small (Did the user like the movie?)\end{tabular}}
          & 1 (liked by the initiator) & \texttt{pkg:likes} \\
          & 0 (disliked by the initiator) & \texttt{pkg:dislikes} \\
          & 2 (not specified) & None \\
        \midrule
        \multirow{3}{*}{\begin{tabular}[c]{@{}c@{}}\texttt{seen} \\ \small (Has the user seen the movie?)\end{tabular}}
          & 1 (seen by the initiator) &\texttt{pkg:hasSeen} \\
          & 0 (unseen by the initiator) & \texttt{pkg:hasNotSeen} \\
          & 2 (not specified) & None \\
        \midrule
        \multirow{2}{*}{\begin{tabular}[c]{@{}c@{}}\texttt{suggested} \\ \small (Who introduced the movie?)\end{tabular}}
          & 1 (suggested by the respondent) & \texttt{pkg:wasSuggested} \\
          & 0 (brought up by the initiator) & None \\
        \bottomrule
    \end{tabular}
\end{table}

To facilitate comparison between extracted triples and ground-truth annotations, we construct an augmented version of the dataset in which the metadata-derived relations are replaced by triples produced by the extraction pipeline. 
This augmentation step is used solely for evaluation and downstream experimentation; it is not required for PKG construction itself. 
By maintaining the same data format as the original annotations, the augmented dataset enables a controlled comparison between PKGs built from LLM-extracted relations and those derived from the original human annotations. 
This allows us to measure how extraction quality influences the performance of downstream recommendation models.

\Cref{alg:pkg-construction} details the PKG construction process.

\begin{algorithm}[t]
\small
\caption{User Preference PKG Construction}
\label{alg:pkg-construction}
\SetKwInOut{Input}{Input}\SetKwInOut{Output}{Output}
\Input{Conversation $C$, LLM model $M$, PKG $G$, training shots $S$, schema $\Sigma$}
\Output{Personal Knowledge Graph $G$}

\tcp{Triple extraction from conversation using the LLM}
prompt $\gets$ formatPrompt($C$, $S$, $\Sigma$) \tcp*{See Algorithm \ref{alg:format-prompt}}
response $\gets$ $M$.generate(prompt) \;
$T \gets$ extractTriples(response) \;

\tcp{Align entities to Wikidata IRIs}
$T' \gets \emptyset$ \;
\ForEach{triple $(s, p, o) \in T$}{
    \If{isEntity($o$)}{
        $(metadata) \gets$ extractMetadata($o$) \;
        
        \tcp{Query Wikidata and fuzzy match on metadata}
        candidates $\gets$ queryWikidata($metadata$) \;
        bestMatch $\gets$ fuzzyMatchBestCandidate(candidates, $metadata$) \;
        
        \If{bestMatch.confidence $\geq$ threshold}{
            $t' \gets$ replaceWithIRI($(s, p, o)$, $o$, bestMatch.iri) \;
            $T' \gets T' \cup \{t'\}$ \;
        }
    }
    \Else{
        $T' \gets T' \cup \{(s, p, o)\}$ \;
    }
}

\tcp{Update the already initialized PKG with new triples}
$G \gets$ addToPKG($T', G$) \;
\Return{$G$}
\end{algorithm}

\begin{algorithm}[htb]
\small
\caption{Format Few-Shot Prompt}
\label{alg:format-prompt}
\SetKwInOut{Input}{Input}\SetKwInOut{Output}{Output}
\Input{Conversation $C$, training shots $S$, schema $\Sigma$}
\Output{Formatted prompt string}

\tcp{Build few-shot examples from training data}
examplePrompts $\gets \emptyset$ \;
\ForEach{shot $\in S$}{
    conversation $\gets$ convertToDialogue(shot) \;
    groundTruth $\gets$ extractGroundTruthTriples(shot) \;
    examplePrompts $\gets$ examplePrompts $\cup$ \{(conversation, groundTruth)\} \;
}

\tcp{Format target conversation}
targetConversation $\gets$ convertToDialogue($C$) \;

\tcp{Construct full prompt with schema and examples}
prompt $\gets$ buildPrompt($\Sigma$, examplePrompts, targetConversation) \;

\Return{prompt}
\end{algorithm}

\subsection{PKG Schema and Semantic Modeling}
We build our PKG using open and interoperable methodologies to ensure that the extracted user-item signals can plug cleanly into larger ecosystems like Wikidata. All the extracted concepts and relations are grounded in RDF and structured through an explicit preference ontology that captures common interaction signals between users and recommended items.

The ontology defines a small set of user–item relations (e.g., preference, interaction history, and recommendation events) that can be consistently represented across conversational domains. While the experimental evaluation in this work focuses on movie recommendation dialogues, the schema itself is not tied to any specific dataset or domain. Instead, it provides a general structure for representing conversationally extracted preferences in a PKG.

\paragraph{Entity Representation with IRIs:}
Each graph node is assigned a stable, dereferenceable IRI for long-term reuse with external KGs.
Users are represented using dataset-provided identifiers, which typically correspond to anonymized participant IDs or session identifiers in conversational datasets. These identifiers uniquely distinguish individual users while preserving privacy, and are incorporated into the PKG namespace using the pattern \texttt{pkg:User\_id}. 
For example, a user with identifier \texttt{42} would be represented as \texttt{pkg:User\_42}.
Entities referenced in conversations are linked to globally recognized identifiers when possible. 

\begin{figure}[h]
    \centering
    \resizebox{\columnwidth}{!}{%
        \begin{tikzpicture}[
  package/.style={rectangle, draw=purple!70, fill=purple!20, rounded corners, minimum width=3.5cm, minimum height=1.5cm, font=\LARGE\bfseries},
  instance/.style={rectangle, draw=blue!70, fill=blue!20, rounded corners, minimum width=3.5cm, minimum height=1.5cm, font=\LARGE\bfseries},
  movie/.style={ellipse, draw=green!70, fill=green!20, minimum width=3.5cm, minimum height=1.5cm, font=\LARGE\bfseries},
  arrow/.style={->, ultra thick},
  edge_label/.style={font=\LARGE, fill=white, rounded corners=3pt, inner sep=4pt}
]

\node[package] (PkgUser) at (0,8) {pkg:User};
\node[instance, right=4cm of PkgUser] (User) {User\_1};
\node[package] (PkgMovie) at (2,2) {pkg:Movie};

\node[movie, right=6cm of User, yshift=4cm]  (Q1)     {wd:Q59687};
\node[movie, right=6cm of User, yshift=1.6cm]  (Q2)   {wd:Q59531};
\node[movie, right=6cm of User, yshift=-1.6cm] (Q3)   {wd:Q59653};
\node[movie, right=6cm of User, yshift=-4.8cm] (Q4)   {wd:Q59317};
\node[movie, right=6cm of User, yshift=-7.6cm] (Q5)   {wd:Q59391};

\node[movie, right=3.5cm of Q1] (L1) {A Touch of Spice};
\node[movie, right=3.5cm of Q2] (L2) {Anatomy of Hell};
\node[movie, right=3.5cm of Q3] (L3) {Argo};
\node[movie, right=3.5cm of Q4] (L4) {Arsenic and Old Lace};
\node[movie, right=3.5cm of Q5] (L5) {Barbara Frietchie};

\path[arrow, dashed] (PkgUser) edge[out=-25, in=180] (User);

\path[arrow] (User) edge[out=60, in=180]  node[edge_label, pos=0.3, yshift=10pt]  {pkg:hasSeen} (Q1);
\path[arrow] (User) edge[out=20, in=180]  node[edge_label, pos=0.5, yshift=7pt]   {pkg:hasNotSeen} (Q2);
\path[arrow] (User) edge[out=-10, in=180] node[edge_label, pos=0.45, yshift=-7pt] {pkg:wasSuggested} (Q3);
\path[arrow] (User) edge[out=-70, in=180] node[edge_label, pos=0.40, yshift=-10pt] {pkg:likes} (Q4);
\path[arrow] (User) edge[out=-95, in=180] node[edge_label, pos=0.40, yshift=-14pt] {pkg:dislikes} (Q5);

\path[arrow] (Q1) edge[out=0, in=180] node[edge_label, pos=0.5, yshift=8pt] {rdfs:label} (L1);
\path[arrow] (Q2) edge[out=0, in=180] node[edge_label, pos=0.5, yshift=8pt] {rdfs:label} (L2);
\path[arrow] (Q3) edge[out=0, in=180] node[edge_label, pos=0.5, yshift=8pt] {rdfs:label} (L3);
\path[arrow] (Q4) edge[out=0, in=180] node[edge_label, pos=0.5, yshift=8pt] {rdfs:label} (L4);
\path[arrow] (Q5) edge[out=0, in=180] node[edge_label, pos=0.5, yshift=8pt] {rdfs:label} (L5);

\path[arrow, dashed] (PkgMovie.east) edge[out=-20, in=180] (Q5);
\path[arrow, dashed] (PkgMovie.east) edge[out=-10, in=180] (Q4);
\path[arrow, dashed] (PkgMovie.east) edge[out=0, in=180]  (Q3);
\path[arrow, dashed] (PkgMovie.east) edge[out=10, in=180]  (Q2);
\path[arrow, dashed] (PkgMovie.east) edge[out=20, in=180]  (Q1);

\end{tikzpicture}
    }
    \caption{User Movie Preference Ontology for the ReDial Dataset. Note that solid edges represent object properties and dashed edges represent \texttt{rdf:type} (instantiation).}
    \label{fig:ontology}
\end{figure}
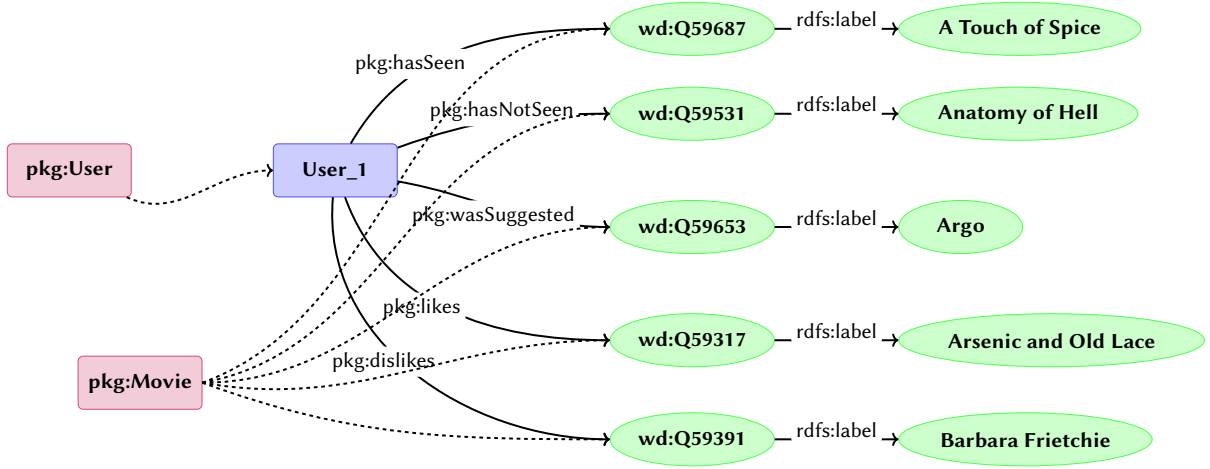

As an example, the ReDial dataset provides worker IDs, which we use in the User namespace as \texttt{pkg:User\_workerId} for user identity.
We also align the extracted movie titles with the corresponding QIDs from the movies' official Wikidata IRIs. For each extracted triple, the referenced movie name is matched against a curated Wikidata index, and the correct QID is inserted into the PKG to enable global resolution and semantic reuse. Using a conversation that indicated a user likes the movie \textit{``The Godfather''} as an example, the extracted triple would be as follows (where \texttt{wd} is the Wikidata and \texttt{pkg} is the User Preference Ontology we defined as illustrated in \Cref{fig:ontology}):
\texttt{(pkg:User\_42, pkg:likes, wd:Q47703)}.
Using the canonical Wikidata identifiers increases interoperability while preserving traceability between extracted triples and the global entity graph.

\paragraph{Ontology Design:}

Our User Movie Preference Ontology (\Cref{fig:ontology}) defines two core entity types:
\begin{itemize}
\item \texttt{pkg:User} — the source of preference and interaction signals
\item \texttt{pkg:Movie} — each anchored to a unique Wikidata entity
\end{itemize}

These are linked by five object properties that capture taste, experience, and conversational recommendation flow:
\begin{itemize}
\item \texttt{pkg:likes}, \texttt{pkg:dislikes} — explicit sentiment
\item \texttt{pkg:hasSeen}, \texttt{pkg:hasNotSeen} — watch status, even when implied
\item \texttt{pkg:wasSuggested} — tracks which movie was recommended to the user
\end{itemize}

All properties use strict domain/range definitions, ensuring only valid user-to-movie edges are created.
This type-safe design prevents malformed triples from entering the graph during extraction or recommendation inference.

\paragraph{Feature Mapping:}
The predicates in the PKG ontology correspond to common user–item interaction signals observed in conversational recommendation settings, such as expressed preference, prior experience, or recommendation events. 
For evaluation purposes, these relations can be mapped to structured annotations available in a given dataset. 
In our experiments, we instantiate this mapping using the metadata provided in the ReDial dataset (\Cref{tab:feature-mapping}), enabling a direct comparison between ground-truth annotations and the triples extracted by the LLM while preserving the intended semantic meaning when serializing the relations into RDF.

\subsection{Query and Usage Patterns}

Once constructed, PKGs can then be queried using SPARQL and used for downstream tasks. Because the PKG is represented in RDF and linked to external knowledge graphs (e.g., Wikidata),  it can support flexible subgraph retrieval and reasoning for personalized services.
One common usage pattern is context retrieval for recommendation systems. A user's PKG may contain preferences across multiple domains (e.g., movies, food, books). For a given task, however, only a relevant subgraph may be needed. For instance, when generating movie recommendations, the system can query only the subset of triples describing the user's interactions with movies, ignoring unrelated preference signals. This enables recommendation models to operate on focused semantic context rather than the entire graph.

The PKG can also evolve dynamically as users continue interacting with conversational systems. New dialogues with an LLM can be processed by the extraction pipeline to generate additional preference triples, incrementally expanding the user's PKG and improving personalization over time.
As an example, the SPARQL query in \Cref{lst:query} retrieves movies that were suggested to a user but have not yet been seen by that user.

\begin{lstlisting}[style=codeblock,label=lst:query, caption={Query Example}]
SELECT ?movie WHERE {
    BIND(pkg:User_42 AS ?user)
    ?user pkg:suggested ?movie . 
    FILTER NOT EXISTS { ?user pkg:seen ?movie . } 
}
\end{lstlisting}

\section{Experimental Design}
We evaluate our triple extraction pipeline using a benchmark conversational recommendation dataset. 
Each dialogue is processed by the extraction model to generate structured $(\texttt{User}, \textit{relation}, \texttt{Item})$ triples. 
These extracted triples are then incorporated into an augmented version of the original dataset, replacing the reference interaction annotations while preserving the underlying conversational content.

This setup enables a controlled comparison between knowledge graphs constructed from LLM-extracted triples and those derived from the original annotations. 
The two variants are evaluated using standard information extraction metrics as well as their impact on downstream recommendation performance.

\subsection{Evaluation Metrics}
To evaluate the quality of the extracted triples, we compare the set of triples generated by the LLM against the reference triples derived from dataset annotations. 
Extraction performance is measured using standard information extraction metrics based on the counts of:
\begin{itemize}
    \item \emph{True Positives (TP)}: Extracted triples that match a ground-truth triple.
    \item \emph{False Positives (FP)}: Extracted triples not found in the ground-truth set.
    \item \emph{False Negatives (FN)}: Ground-truth triples not extracted by the model.
\end{itemize}

A triple is considered a match if the subject, predicate, and object all correspond to the same entities and relations after normalization and entity linking.

From these counts, we calculate $precision = \frac{TP}{TP + FP}$, $recall= \frac{TP}{TP + FN}$, and $F1= 2 \cdot \frac{precision \cdot recall}{precision + recall}$. 

Precision reflects the correctness of the model’s predictions, while recall captures their completeness.
The F1-score summarizes overall extraction performance by balancing these two aspects.
Metrics are computed over the set of triples extracted from each dialogue and aggregated across the dataset.

\subsection{Model and Prompting Setup}

To evaluate the robustness of the proposed extraction pipeline across model architectures and scales, we conduct experiments using models from the Qwen3~\cite{huggingface2024qwen3} and Gemma-3 Instruct~\cite{google2025gemma3} families. Using our prompt template (available in our repository\footnote{\url{https://github.com/brains-group/LLMTripleExtractor/blob/master/prompt_template.md}}), we perform ablation experiments across zero to ten-shot prompting settings. Few-shot examples follow our shot template\footnote{\url{https://github.com/brains-group/LLMTripleExtractor/blob/master/shot_example.md}} and are prepended to the target conversation during inference. To study the effect of model scale, we evaluate several representative sizes from each family. Because the two families do not provide identical parameter counts, we select the closest available sizes to enable a comparable evaluation across scales. The specific model variants used in our experiments are shown in \Cref{fig:F1Aug}. 
 
Although the Qwen3 models are base language models and the Gemma-3 models are instruction-tuned, the objective of this study is not to perform a head-to-head benchmark between model families. Rather, we use these models to examine how extraction performance varies with model scale, alignment, and prompting strategy within the proposed PKG construction pipeline.

\subsection{Downstream Task Evaluation}
To assess the practical utility of the extracted triples, we evaluate their impact on a downstream recommendation task. 
We run triple extraction on the full ReDial dataset using the models described above and obtain a set of extracted triples per run. Using these triples, we augment the original dataset to obtain the same number of augmented datasets for evaluation. To evaluate the impact of extracted triples on recommendation performance, we compare models trained on the augmented ReDial dataset against models evaluated on the original ground-truth ReDial triples. 

Specifically, we evaluate the effectiveness of the downstream task on the FedTREK-LM recommendation system~\cite{spadea2026federated,fedtreklmgithub}. This recommendation system uses a local LLM in a federated system to generate recommendations using PKGs for context, so it fits nicely with our local, LLM-based, triple-extraction pipeline. 
This framework constructs PKGs from the ReDial dataset and trains on the conversations to learn good recommendations, followed by evaluating itself on a separate test set from the dataset. To evaluate our augmented dataset in this framework, we construct the FedTREK-LM dataset with our augmented dataset as the base, but then we still use the original ground-truth test dataset for evaluation. This way, we can compare the results with the original ground-truth training dataset against the results with our augmented dataset constructed from our extracted triples. For our primary comparison metric, we will look at the F1-scores of the recommendation models.
This experiment, therefore, measures whether the triples extracted by the LLM are sufficiently accurate and semantically meaningful to support downstream personalization tasks.

\section{Results}
\begin{figure}[htb]
    \centering
    \includegraphics[width=0.9\linewidth]{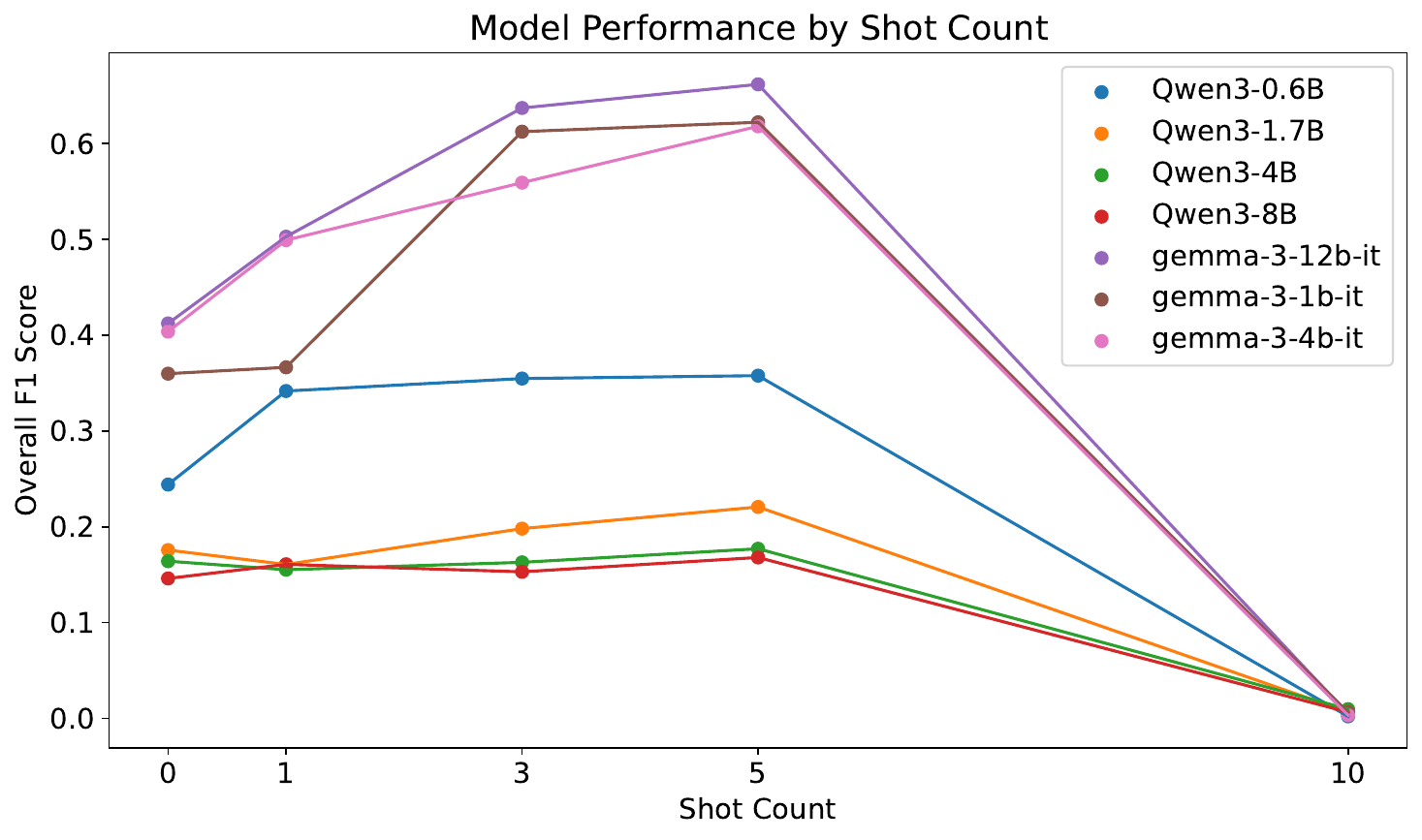}
    \caption{F1 Scores Across Model Sizes}
    \label{fig:F1Aug}
\end{figure}

In this section, we primarily analyze the impact of model size and family on these metrics with a brief look into the effect of k-shot prompting. A comprehensive breakdown of the triple extraction performance across all tested models, families, and prompting configurations is provided in \Cref{tab:evaluation_metrics}. This table details the precision, recall, and F1-scores for individual relation categories as well as the macro-averaged overall metrics, providing a granular view of the extraction fidelity. Our best performing model for extraction was Gemma-12B Instruct (as shown in \Cref{fig:F1Aug}), with the Gemma-3 Instruct family outperforming Qwen3. All models showed an increase in performance until a certain shot setting, with a stark drop at the 10-shot setting. Performance generally scaled with the number of shots until a significant decline at the 10-shot mark. This suggests a threshold where the models begin to overfit to the provided examples, prioritizing the prompt's fixed patterns over the unique conversational structures in the test data. As our testing jumped from 5 to 10 shots, an optimal inflection point likely exists within that interval before performance degrades.

\begin{table*}[htb!]
\centering
\caption{Triple extraction performance metrics (Precision, Recall, and F1-score) broken down by model, shot count (S), and relation category. The highest score for each model across shot settings is in \textbf{bold} font, while the global maximum for each column is marked with an upward arrow ($\uparrow$).}
\setlength{\tabcolsep}{2pt}
\resizebox{\textwidth}{!}{
\begin{tabular}{lc|ccc|ccc|ccc|ccc}
\toprule
\multirow{2}{*}{\textbf{Model}} & \multirow{2}{*}{\textbf{S}} & \multicolumn{3}{c|}{\textbf{Liked}} & \multicolumn{3}{c|}{\textbf{Seen}} & \multicolumn{3}{c|}{\textbf{Suggested}} & \multicolumn{3}{c}{\textbf{Overall}} \\
 & & Prec & Rec & F1 & Prec & Rec & F1 & Prec & Rec & F1 & Prec & Rec & F1 \\
\midrule
\multirow{5}{*}{\begin{tabular}[c]{@{}l@{}}Qwen3\\(0.6B)\end{tabular}} & 0 & \textbf{0.9014} & 0.2088 & 0.3391 & \textbf{0.8422} & 0.1091 & 0.1932 & \textbf{0.9281} & 0.1134 & 0.2021 & \textbf{0.8926} & 0.1414 & 0.2441 \\
 & 1 & 0.8827 & 0.3277 & 0.4780 & 0.7420 & \textbf{0.1615} & \textbf{0.2653} & 0.8790 & 0.1662 & 0.2795 & 0.8416 & 0.2144 & 0.3417 \\
 & 3 & 0.8758 & 0.3783 & 0.5284 & 0.7355 & 0.1402 & 0.2355 & 0.8509 & \textbf{0.1723} & \textbf{0.2866} & 0.8359 & 0.2251 & 0.3547 \\
 & 5 & 0.8726 & \textbf{0.3997} & \textbf{0.5483} & 0.6828 & 0.1520 & 0.2486 & 0.8558 & 0.1535 & 0.2603 & 0.8183 & \textbf{0.2289} & \textbf{0.3577} \\
 & 10 & 0.7785 & 0.0024 & 0.0048 & 0.3418 & 0.0010 & 0.0020 & 0.0000 & 0.0000 & 0.0000 & 0.5601 & 0.0011 & 0.0022 \\
\midrule
\multirow{5}{*}{\begin{tabular}[c]{@{}l@{}}Qwen3\\(1.7B)\end{tabular}} & 0 & 0.9267 & 0.1641 & 0.2788 & \textbf{0.8309} & 0.1057 & 0.1875 & \textbf{0.9750} & 0.0323 & 0.0625 & \textbf{0.8946} & 0.0974 & 0.1757 \\
 & 1 & \textbf{0.9281} & 0.1277 & 0.2245 & 0.8285 & 0.0890 & 0.1607 & 0.9318 & 0.0538 & 0.1017 & 0.8928 & 0.0883 & 0.1607 \\
 & 3 & 0.9172 & 0.1647 & 0.2793 & 0.8168 & 0.0980 & 0.1750 & 0.9199 & 0.0784 & 0.1445 & 0.8860 & 0.1115 & 0.1981 \\
 & 5 & 0.9088 & \textbf{0.1863} & \textbf{0.3092} & 0.8217 & \textbf{0.1097} & \textbf{0.1936} & 0.9317 & \textbf{0.0894} & \textbf{0.1631} & 0.8870 & \textbf{0.1260} & \textbf{0.2207} \\
 & 10 & 0.8333 & 0.0082 & 0.0162 & 0.6408 & 0.0017 & 0.0034 & 0.7377 & 0.0008 & 0.0016 & 0.7864 & 0.0034 & 0.0068 \\
\midrule
\multirow{5}{*}{\begin{tabular}[c]{@{}l@{}}Qwen3\\(4B)\end{tabular}} & 0 & \textbf{0.9623}$\uparrow$ & \textbf{0.1526} & \textbf{0.2634} & 0.9310 & 0.1050 & 0.1887 & \textbf{0.9922}$\uparrow$ & 0.0216 & 0.0423 & \textbf{0.9523} & 0.0898 & 0.1641 \\
 & 1 & 0.9569 & 0.1237 & 0.2191 & \textbf{0.9379} & 0.0959 & 0.1740 & 0.9726 & 0.0401 & 0.0770 & 0.9522 & 0.0845 & 0.1552 \\
 & 3 & 0.9587 & 0.1211 & 0.2150 & 0.9299 & 0.1001 & 0.1807 & 0.9753 & 0.0514 & 0.0977 & 0.9510 & 0.0891 & 0.1629 \\
 & 5 & 0.9585 & 0.1279 & 0.2257 & 0.9231 & \textbf{0.1118} & \textbf{0.1994} & 0.9722 & \textbf{0.0585} & \textbf{0.1104} & 0.9475 & \textbf{0.0977} & \textbf{0.1771} \\
 & 10 & 0.8189 & 0.0059 & 0.0117 & 0.6811 & 0.0046 & 0.0091 & 0.6757 & 0.0042 & 0.0083 & 0.7252 & 0.0049 & 0.0097 \\
\midrule
\multirow{5}{*}{\begin{tabular}[c]{@{}l@{}}Qwen3\\(8B)\end{tabular}} & 0 & \textbf{0.9618} & 0.1372 & 0.2401 & 0.9332 & 0.0863 & 0.1580 & \textbf{0.9902} & 0.0222 & 0.0434 & 0.9539 & 0.0791 & 0.1461 \\
 & 1 & 0.9596 & \textbf{0.1402} & \textbf{0.2447} & \textbf{0.9403}$\uparrow$ & 0.0958 & 0.1739 & 0.9767 & 0.0354 & 0.0683 & \textbf{0.9549}$\uparrow$ & 0.0878 & 0.1608 \\
 & 3 & 0.9559 & 0.1183 & 0.2105 & 0.9367 & 0.0982 & 0.1778 & 0.9772 & 0.0391 & 0.0752 & 0.9517 & 0.0832 & 0.1530 \\
 & 5 & 0.9579 & 0.1232 & 0.2183 & 0.9387 & \textbf{0.1122} & \textbf{0.2004} & 0.9754 & \textbf{0.0468} & \textbf{0.0893} & 0.9531 & \textbf{0.0921} & \textbf{0.1680} \\
 & 10 & 0.8285 & 0.0050 & 0.0099 & 0.6811 & 0.0046 & 0.0091 & 0.7377 & 0.0008 & 0.0016 & 0.7473 & 0.0033 & 0.0066 \\
\midrule
\multirow{5}{*}{\begin{tabular}[c]{@{}l@{}}gemma-3\\(1b-it)\end{tabular}} & 0 & 0.8208 & \textbf{0.7243} & \textbf{0.7695} & \textbf{0.7340} & 0.0121 & 0.0238 & \textbf{0.9580} & 0.0089 & 0.0176 & \textbf{0.8207} & 0.2304 & 0.3598 \\
 & 1 & \textbf{0.8495} & 0.4069 & 0.5502 & 0.6578 & 0.2438 & 0.3557 & 0.9086 & 0.0916 & 0.1664 & 0.7793 & 0.2395 & 0.3664 \\
 & 3 & 0.8363 & 0.6902 & 0.7563 & 0.6369 & 0.4876 & 0.5523 & 0.7823 & 0.3983 & 0.5279 & 0.7485 & 0.5180 & 0.6123 \\
 & 5 & 0.8339 & 0.7077 & 0.7656 & 0.6194 & \textbf{0.5092} & \textbf{0.5589} & 0.7722 & \textbf{0.4197}$\uparrow$ & \textbf{0.5438} & 0.7369 & \textbf{0.5383}$\uparrow$ & \textbf{0.6221} \\
 & 10 & 0.8369 & 0.0038 & 0.0076 & 0.7167 & 0.0030 & 0.0060 & 0.6695 & 0.0026 & 0.0052 & 0.7411 & 0.0031 & 0.0062 \\
\midrule
\multirow{5}{*}{\begin{tabular}[c]{@{}l@{}}gemma-3\\(4b-it)\end{tabular}} & 0 & \textbf{0.8914} & 0.5752 & 0.6992 & 0.8265 & 0.2478 & 0.3813 & \textbf{0.9439} & 0.0080 & 0.0159 & \textbf{0.8704} & 0.2628 & 0.4037 \\
 & 1 & 0.8745 & 0.6729 & 0.7606 & \textbf{0.8284} & 0.3892 & 0.5296 & 0.9220 & 0.0407 & 0.0780 & 0.8587 & 0.3518 & 0.4991 \\
 & 3 & 0.8605 & 0.7258 & 0.7874 & 0.8221 & 0.4092 & 0.5464 & 0.9070 & 0.1551 & 0.2649 & 0.8532 & 0.4158 & 0.5591 \\
 & 5 & 0.8445 & \textbf{0.7902}$\uparrow$ & \textbf{0.8164}$\uparrow$ & 0.8048 & \textbf{0.5316} & \textbf{0.6403} & 0.9033 & \textbf{0.1918} & \textbf{0.3164} & 0.8372 & \textbf{0.4896} & \textbf{0.6179} \\
 & 10 & 0.8041 & 0.0023 & 0.0046 & 0.7167 & 0.0030 & 0.0060 & 0.0000 & 0.0000 & 0.0000 & 0.7507 & 0.0017 & 0.0034 \\
\midrule
\multirow{5}{*}{\begin{tabular}[c]{@{}l@{}}gemma-3\\(12b-it)\end{tabular}} & 0 & \textbf{0.9341} & 0.4347 & 0.5933 & 0.8747 & 0.3097 & 0.4574 & \textbf{0.9460} & 0.0808 & 0.1489 & 0.9114 & 0.2663 & 0.4122 \\
 & 1 & 0.9340 & 0.4690 & 0.6244 & \textbf{0.8779} & 0.4062 & 0.5554 & 0.9449 & 0.1863 & 0.3112 & \textbf{0.9132} & 0.3468 & 0.5027 \\
 & 3 & 0.9237 & 0.5595 & 0.6969 & 0.8767 & 0.5326 & 0.6626 & 0.9262 & \textbf{0.3930} & \textbf{0.5518}$\uparrow$ & 0.9068 & 0.4909 & 0.6370 \\
 & 5 & 0.9062 & \textbf{0.6451} & \textbf{0.7537} & 0.8747 & \textbf{0.6082}$\uparrow$ & \textbf{0.7175}$\uparrow$ & 0.9428 & 0.3376 & 0.4972 & 0.9017 & \textbf{0.5226} & \textbf{0.6617}$\uparrow$ \\
 & 10 & 0.8369 & 0.0038 & 0.0076 & 0.6615 & 0.0031 & 0.0062 & 0.4583 & 0.0002 & 0.0004 & 0.7315 & 0.0023 & 0.0046 \\
\bottomrule
\end{tabular}
}
\label{tab:evaluation_metrics}
\end{table*}

\begin{figure}[htb]
    \centering
    \includegraphics[width=\linewidth]{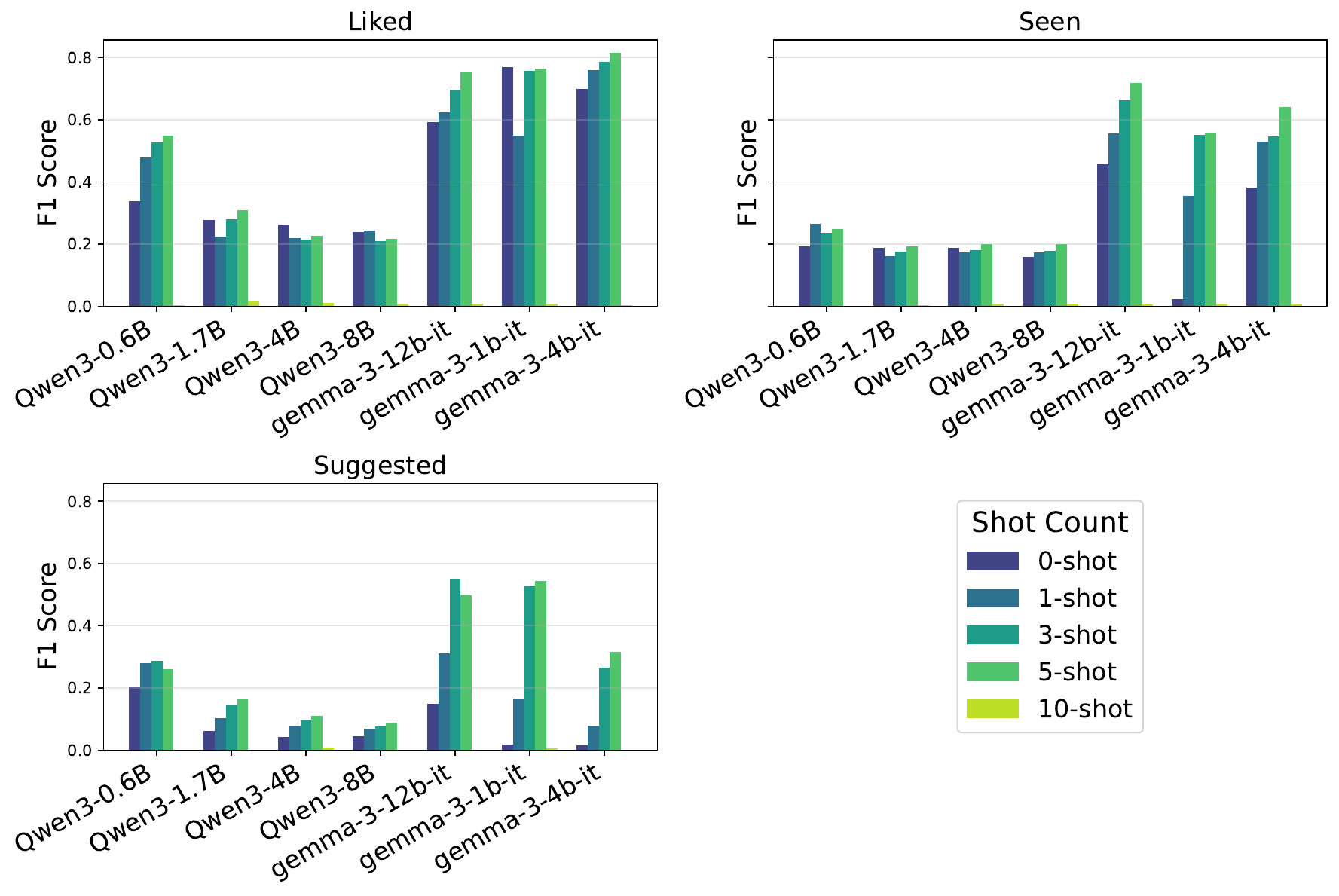}
    \caption{F1 Scores Across Model Relation}
    \label{fig:F1Rel}
\end{figure}
\paragraph{Which relations tell the strongest signal?}
In \Cref{fig:F1Rel}, the clearest patterns emerged when the data was tied directly to user taste:
\begin{itemize}
\item \texttt{liked} has the highest F1-score across the board, making it the most reliable source for recommendations. It captures enthusiasm and critique well, even when negative cases are included.
\item \texttt{seen} is solid but softer. The model can detect it, but watch status is often implied, and implication is harder for LLMs to pin down.
\item \texttt{suggested} is the trickiest. Tracing who introduced which movie sometimes requires long-range conversational memory. Unsurprisingly, smaller models struggle the most here, with particularly low F1-scores. However,  this has less impact on the final recommendation goal than \texttt{liked}.
\end{itemize}

Although a granular analysis of individual triple mistmatches was not performed, a systematic analysis of the relation-specific metrics suggests that the observed recall gaps are rooted in the complexity of conversation that relations appear in. The significantly higher F1 scores for the ``liked'' relation across all models indicate that preference sentiment is typically expressed through explicit, high signal tokens that were easily captured during extraction. In contrast, relations with lower recall reflect the inherent ambiguity of conversational context. Additionally, the performance disparity can be attributed to the fact that the Gemma models were instruction-tuned, which may play a role in ambiguity resolution.

\subsection{Downstream Task Results}

\begin{figure}[htb]
    \centering
    \includegraphics[width=\textwidth]{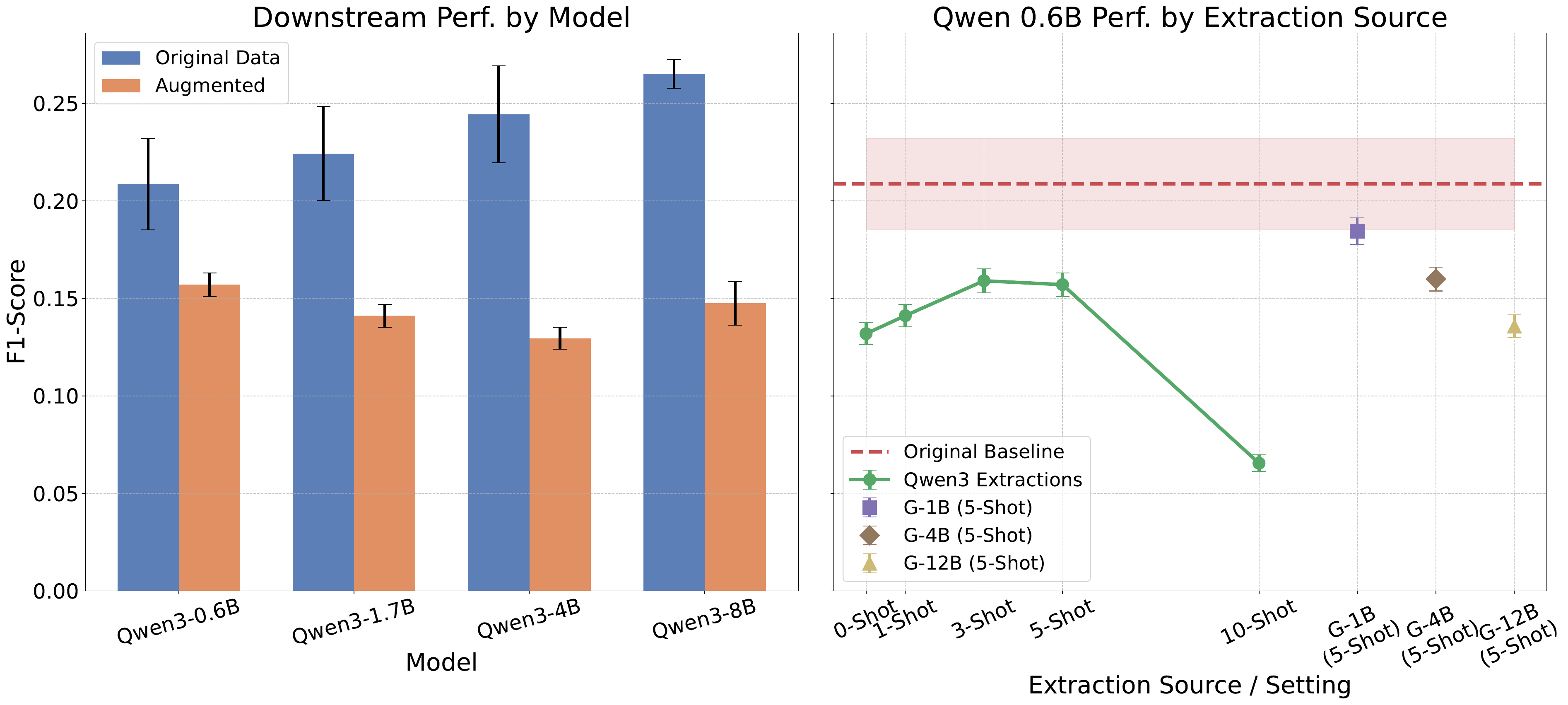}
    \caption{Downstream recommendation performance evaluated on the FedTREK-LM framework. (Left) Comparison of F1-scores across Qwen3 model sizes using the original ReDial dataset versus the 5-shot augmented dataset. (Right) Downstream F1-score for the Qwen3-0.6B model across varying extraction sources and settings, comparing Qwen3 extractions from 0 to 10 shots (line) against 5-shot extractions from the Gemma-3 model family (scatter points).}
    \label{fig:downstream_results}
\end{figure}

As outlined in Section 4.3, we evaluate the utility of our extracted triples by training the FedTREK-LM recommendation framework on our augmented datasets and comparing the results against models trained on the original ground-truth ReDial triples. For our baseline augmented datasets, we focus specifically on the 5-shot extractions. As detailed in \Cref{tab:evaluation_metrics}, the 5-shot setting consistently yielded the peak, or near-peak, overall F1-scores during upstream extraction across all evaluated Qwen3 and Gemma-3 model sizes, making it the most stable and representative configuration before the severe overfitting and performance degradation observed at 10 shots. To circumvent VRAM scaling constraints, we restrict the downstream recommendation models to the Qwen3 family.

\Cref{fig:downstream_results} (left) illustrates the recommendation performance across different downstream model sizes. While transitioning from ground-truth annotations to LLM-extracted datasets naturally results in an F1-score degradation, this drop is surprisingly low relative to the upstream triple extraction scores for the smallest model. For example, \Cref{tab:evaluation_metrics} shows that the Qwen3-0.6B model achieved an overall extraction F1-score of only 0.3577 under the 5-shot setting; however, it maintained over 75\% of its ground-truth F1-score baseline (0.1571 vs. 0.2087) in the downstream recommendation task. This resilience can likely be attributed to an implicit filtering effect: successfully extracted triples tend to originate from structurally clear conversations that the downstream LLM can inherently parse more effectively, naturally filtering out noisy or meandering datapoints. However, a stark degradation and general inverse scaling trend emerges with the larger models. While the original ground-truth data shows downstream performance scaling positively with model size, peaking at an F1-score of 0.2652 for Qwen3-8B, the augmented datasets cause the larger models to falter significantly. Despite a marginal improvement over the 4B model, the 8B model's downstream F1-score plummets to 0.1475, failing to surpass even the 0.6B model's augmented performance. This collapse directly correlates with their upstream extraction failures; as shown in \Cref{tab:evaluation_metrics}, the 4B and 8B models struggled with the upstream extraction task, yielding poor overall F1-scores of 0.1771 and 0.1680, primarily driven by extremely low recall. The severely diminished quality of their extracted PKGs forms a sparse, weak contextual foundation that entirely bottlenecks their naturally higher downstream recommendation capabilities.

\Cref{fig:downstream_results} (right) details the impact of prompt scaling and extraction source on downstream utility, specifically utilizing the Qwen3-0.6B downstream model. Looking at the Qwen3 extractions, downstream performance steadily improves as the shot count increases, peaking at 3 shots with an F1-score of 0.159, before experiencing a slight dip at 5 shots and a severe degradation at 10 shots. Cross-referencing this with \Cref{tab:evaluation_metrics} reveals that while 5-shot extraction had a higher overall F1-score, the 3-shot extraction maintained a higher overall precision (0.8359 vs. 0.8183). This performance trade-off suggests a high downstream sensitivity to extraction precision within the same model family, indicating that wrong extractions (false positives) actively harm the recommendation task more than marginal increases in correct extractions help.

To further investigate the impact of extraction quality, we trained the Qwen3-0.6B downstream model using triples extracted by the Gemma-3 models under the optimal 5-shot setting. Intriguingly, we observe an inverse scaling relationship: the extractions from the smallest model, Gemma-1B, yielded the highest downstream F1-score (0.184), significantly outperforming the extractions from the 4B and 12B models. By analyzing the relation-specific metrics in \Cref{tab:evaluation_metrics}, we can attribute this to the 1B model's more holistic and balanced extraction profile. While the 12B model achieved higher overall precision, Gemma-1B maintained a significantly higher recall for the critical "Liked" category (0.7077 vs. 0.6451) and demonstrated much greater consistency in capturing the conversational flow, achieving the highest F1-score for the difficult "Suggested" relation (0.5438). This indicates that for a smaller downstream model like Qwen3-0.6B, a consistently populated, diverse knowledge graph, one that reliably captures explicit preferences alongside conversational context, provides a richer and more effective foundation for learning than a sparse, high-precision graph. Crucially, the fact that optimal downstream utility is achieved using triples extracted by a 1-billion parameter model validates our overarching goal of enabling decentralized, privacy-preserving PKG construction. Because extracting robust preference signals does not strictly require massive, computationally expensive models, this pipeline proves highly viable for efficient, on-device applications.

\section{Conclusion}
We present an evaluation of open-weight LLMs for extracting structured user preferences from recommendation dialogues, with the goal of populating PKGs directly from conversational input. While our upstream experiments showed that the Gemma-3-12B Instruct model achieved the highest overall extraction F1-score, our downstream recommendation analysis revealed a more nuanced reality: smaller, highly balanced models like Gemma-3-1B provide a more effective foundation for downstream learning, validating the viability of efficient, on-device PKG construction. Furthermore, highly capable larger models like Qwen3-8B demonstrated an ability to overcome sparse extractions through superior reasoning capacity. While this work employs standard machine learning evaluation metrics, its core contribution is to the Semantic Web community: we validate a reproducible pipeline for building PKGs from dialogue and analyze how extraction fidelity shapes downstream recommendation workflows.

To address the scalability and generalizability of this approach, we note that while this study utilized the ReDial dataset and its corresponding metadata features as a ground-truth framework for rigorous evaluation, our core extraction methodology is fundamentally dataset- and schema-agnostic. The prompt-based extraction relies on natural language understanding of multi-turn dialogues, meaning it can theoretically be applied to any conversational corpus. The specific ontology mapping used in this work was designed to align with ReDial's structure for benchmarking purposes; however, the pipeline can seamlessly be adapted to target different downstream ontologies or domain schemas without requiring manual metadata mapping in real-world deployments.

This work demonstrates both the feasibility and current limitations of using LLMs to bootstrap PKGs for personalized recommendation systems. In contrast to traditional approaches that rely on explicit, centralized logging or extensive user instrumentation, our method offers a scalable alternative that derives preferences from natural dialogue. These findings open up several promising directions for future work, including better prompt engineering, lightweight fine-tuning, and structured reasoning over extracted triples.

As personalization needs evolve, dynamic PKG adaptation becomes increasingly important. Advances in knowledge editing and continual learning offer mechanisms to insert or update specific facts in PKGs before they are presented to an LLM, altering functionality without retraining. These tools could enable PKGs to not only be constructed from conversations but also updated to reflect changing user preferences, a capability missing from static graph models and current dialogue-based extraction pipelines.

A central challenge in many KG-based personalized service applications such as recommenders is the cold-start problem; obtaining high-quality, explicitly stated structured user preferences is labor-intensive and often infeasible at scale. By contrast, this paper shows that LLMs can effectively extract preference triples from both human-human and human-chatbot conversations, offering a low-friction way to initiate and refine PKGs over time.
Such PKGs can also support retrieval-augmented generation (RAG) pipelines, enabling more transparent, controllable, and personalized responses from LLM-based systems. 
In conclusion, we believe our framework provides a valuable diagnostic lens on the current capabilities of LLMs in extracting user-specific knowledge, a critical step toward adaptive, conversational AI grounded in structured personalization.

\subsection*{Supplementary Material Statement:}
\label{sec:resources}

All research artifacts, including source code, dataset construction scripts, and result generation pipelines, are available in our GitHub repository. All external datasets and software dependencies used in this work are documented and linked in the repository’s README.\\
\url{https://github.com/brains-group/LLMTripleExtractor}. 

\subsection*{GenAI Usage Disclosure:}
GenAI was used to help generate some parts of the code as well as the paper content. The authors retain full responsibility for the accuracy, originality, and overall content of the code and the paper.

\bibliography{references}
\end{document}